\documentclass[aps,pra,showpacs,twocolumn,longbibliography]{revtex4-2}

\usepackage{graphicx}
\usepackage{dcolumn}
\usepackage{bm}
\usepackage{color}
\usepackage{amsmath}
\usepackage{amssymb}
\newcommand{\tabincell}[2]{\begin{tabular}{@{}#1@{}}#2\end{tabular}}
\usepackage[colorlinks,citecolor=red,urlcolor=red,bookmarks=false,hypertexnames=true]{hyperref}

\begin{document}

\title{Mid-infrared spectrally-pure single-photon states generation from 22 nonlinear optical crystals}
\author{Wu-Hao Cai$^{1, 2, 3}$}
\author{Ying Tian$^{1}$}
\author{Rui-Bo Jin$^{1, 4}$}
\email{jrbqyj@gmail.com}

\date{\today }
\affiliation{$^{1}$Hubei Key Laboratory of Optical Information and  Pattern Recognition, Wuhan Institute of Technology, Wuhan 430205, China}

\affiliation{$^{2}$Research Institute of Electrical Communication, Tohoku University, 2-1-1 Katahira, Sendai, 980-8577, Japan}

\affiliation{$^{3}$Graduate School of Engineering, Tohoku University, 6-6 Aramaki Aza Aoba, Aoba-ku, Sendai, 980-8579, Japan}

\affiliation{$^{4}$Guangdong Provincial Key Laboratory of Quantum Science and Engineering, Southern University of Science and Technology, Shenzhen 518055, China}




\begin{abstract}
We theoretically investigate the preparation of pure-state single-photon source from 14 birefringent crystals (CMTC, THI, LiIO$_3$, AAS, HGS, CGA, TAS, AGS, AGSe, GaSe, LIS, LISe, LGS, and LGSe) and 8 periodic poling crystals (LT, LN, KTP, KN, BaTiO$_3$, MgBaF$_4$, PMN-0.38PT, and OP-ZnSe) in a wavelength range from 1224 nm to 11650 nm. The three kinds of group-velocity-matching (GVM) conditions, the phase matching conditions, the spectral purity, and the Hong-Ou-Mandel interference are calculated for each crystal. This study may provide high-quality single-photon sources for quantum sensing, quantum imaging, and quantum communication applications at the mid-infrared wavelength range.
\end{abstract}

\maketitle


\section{Introduction}\label{sec1}
Single-photon source at mid-infrared (MIR) wavelength range (approximately 2--20 $\mu$m) has important potential applications in quantum sensing, quantum imaging, and quantum communication \cite{Ebrahim-Zadeh2008book,Tournie2019book,Walmsley2005,Slussarenko2019APR}.
Firstly, the 3--5 $\mu$m band  contains  absorption peaks of many gases, such as H$_2$O, CO, CO$_2$, SO$_2$, and SO$_3$ \cite{Caffey2011}, so this band is important for sensing of these gases in environmental monitoring \cite{Shamy2020SR, Chen2020OLE}; 
the 7--10 $\mu$m band  contains  absorption peaks of  H$_2$O$_2$, CH$_4$, O$_3$, TAT, Acetone, and sarin \cite{Caffey2011}, etc, therefore this band is important for the sensing chemical or explosive materials in the applications of industrial production or defense security. 
The single-photon source in these gas sensing applications may  provide an ultra-high sensitivity \cite{Hogstedt2014}.
Secondly, 3--5 $\mu$m and 8--14 $\mu$m are two widely used ranges for MIR thermal infrared imaging camera for medical and forensic usage \cite {Edelman2013}, since the room temperature objects emit light at these wavelength ranges. 
And MIR single-photon sources can provide a diagnosis in a non-invasive manner, which is important for medical or biological samples \cite{Shi2019NP}.
Thirdly, 3--5 $\mu$m is also an atmospheric transmission window with relatively high transparency, which is useful for large-scale free-space quantum communications, such as entanglement distribution \cite{Yin2017}, quantum key distribution \cite{Wang2022}, or quantum direct communication \cite{Zhou2020}.

Spontaneous parametric down-conversion (SPDC) and four-wave mixing (FWM) are two widely used methods to prepare single photons. 
Many previous works have been dedicated to the development of high-quality single-photon source or entangled photon source in the MIR range from an SPDC or FWM process.
On the experimental side, PPLN \cite{Mancinelli2017NC, Sua2017SR, Arahata2021}, GaP \cite{Kuo2019CLEO}, and silicon waveguide \cite{Rosenfeld2020} have been investigated to prepare single-photon source; PPLN \cite{Prabhakar2020SA} has been studied for entangled photon source generation. 
On the theory side, PPLN \cite{Hojo2021}, PPKN \cite{Lee2016AO}, PMN-0.38PT \cite{Kundys2020AQT}, etc have been investigated for single-photon source \cite{ McCracken2018JOSAB, Wei2021, Cai2022, Zhu2023JOSAB, Zhang2022Wulixuebao}; p-doped semiconductor \cite{Razali2016SST} has been studied for entangled photon source.  
In addition, some studies explored single-photon detection by superconducting nanowire single-photon detector (SNSPD)  \cite{Marsili2012, Bellei2016OE, Kuo2018OSAC} or by silicon avalanche photodiode (SAPD) in an up-conversion configuration \cite{Mancinelli2017NC, Piccione2018SPIE}.

However, the previous studies are still insufficient for the need of MIR applications.
On one hand, the previous experimental work is mainly focused on PPLN crystal and the wavelength range is below 5 $\mu$m. SO, the range from 5--20 $\mu$m is still needs further exploration.
On the other hand, the spectrally-pure single photon source is proved to be a good resource \cite{Grice2001PRA}, but this source is still rare  because the group-velocity matching (GVM) conditions can only be matched at very limited wavelengths in a crystal.
Therefore, it is still necessary to explore more nonlinear optical crystals to fully meet the need of MIR band applications. For this purpose, in this work, we investigate MIR spectrally-pure single-photon generation from 14 crystals by the birefringence phase matching (BPM) method, and 8 crystals by the quasi-phase-matching (QPM) method. They can meet three kinds of GVM conditions and prepare spectrally uncorrelated biphotons so as to generate spectrally pure heralded  single-photon state.

\begin{table*}[tbp]
\centering\begin{tabular}{c|ccccc}
\hline      \hline
Name (Ref.)    &chemical formula             & axis      & point group & $\lambda_{transp.}$ ($\mu$m) & $d_{max(wavelength)}$ (pm/V)     \\
 \hline
CMTC    &$\mathrm{CdHg(SCN)_4}$       & uniaxial  &$\bar4$      &  0.40$\sim$2.35              &$d_{31(1.064)}$ = 6.2 $\pm$ 1.2  \\
 \hline
 THI    &$\mathrm{Tl_4HgI_6}$          & uniaxial & 4mm         &  1.00$\sim$60.0              & unknown  \\
 \hline
LiIO$_3$ \cite{Kato1985IEEE}   &$\mathrm{LiIO_3}$         & uniaxial  & 6           &  0.28$\sim$6.00              & $d_{33(1.064)}$ = 4.6 $\pm$ 0.3 \\
 \hline
AAS     &$\mathrm{Ag_3AsS_3}$         & uniaxial  & 3m          &  0.61$\sim$13.3              & $d_{22(1.064)}$ = 16.6 $\pm$ 2.5\\
 \hline
HGS \cite{Kato2016AO}    &$\mathrm{HgGa_2S_4}$         & uniaxial  &$\bar4$      &  0.55$\sim$13.0              & $d_{36(1.064)}$ = 31.5 $\pm$ 4.7\\
 \hline
CGA \cite{Bhar1976AO}    &$\mathrm{CdGeAs_2}$          & uniaxial  &$\bar4$2m    &  2.30$\sim$18.0              & $d_{36(10.6)}$ = 186 $\pm$ 16   \\
 \hline
TAS     &$\mathrm{Tl_3AsSe_3}$        & uniaxial  & 3m          &  1.28$\sim$17.0              & $d_{+(10.6)}$ = 68 $\pm$ 31   \\
 \hline
AGS     &$\mathrm{AgGaS_2}$           & uniaxial  &$\bar4$2m    &  0.47$\sim$13.0              & $d_{36(10.6)}$ = 12.5 $\pm$ 2.5  \\
 \hline
AGSe    &$\mathrm{AgGaSe_2}$          & uniaxial  &$\bar4$2m    &  0.71$\sim$19.0              & $d_{36(10.591)}$ = 39.5 $\pm$ 1.9 \\
 \hline
GaSe \cite{Kato2013}   &$\mathrm{GaSe}$              & uniaxial  &$\bar6$2m    &  0.62$\sim$20.0              & $d_{22(10.6)}$ = 54 $\pm$ 11 \cite{Kato2013}   \\
 \hline\hline
LIS  \cite{Fossier2004JOSAB}   &$\mathrm{LiInS_2}$           & biaxial   & mm2         &  0.34$\sim$13.2              & $d_{33(2.3)}$ = $-$16 $\pm$ 4   \\
 \hline
LISe  \cite{Kato2014}   &$\mathrm{LiInSe_2}$          & biaxial   & mm2         &  0.46$\sim$14.0              & $d_{31(2.3)}$ = $-$16  $\pm$ 4 \cite{Petrov2010}\\
 \hline
LGS \cite{Kato2017LGS}   &$\mathrm{LiGaS_2}$            & biaxial   & mm2         &  0.32$\sim$11.6              & $d_{33(2.3)}$ = $-$10.7 $\pm$ 2.7 \cite{Petrov2004}\\
 \hline
LGSe  \cite{Miyata2017}  &$\mathrm{LiGaSe_2}$           & biaxial   & mm2         &  0.37$\sim$13.2              & $d_{33(2.3)}$ = $-$18.2 $\pm$ 4.6 \cite{Petrov2004}\\
 \hline
 \hline
LT  \cite{Dolev2009}   &$\mathrm{LiTaO_3}$         & uniaxial  & 3m          & 0.28$\sim$5.50               & $d_{33(1.064)}$ = 12.9 \\
 \hline
LN     &$\mathrm{LiNbO_3}$            & uniaxial  & 3m          & 0.40$\sim$5.50               & $d_{33(1.064)}$ = 25.2 \\
 \hline
KTP    &$\mathrm{KTiOPO_4}$           & biaxial   & mm2         & 0.35$\sim$4.50               & $d_{33(1.064)}$ = 14.6 $\pm$ 0.7\\
 \hline
KN     &$\mathrm{KNbO_3}$            & biaxial   & mm2          & 0.40$\sim$4.50               & $d_{11(1.064)}$ = 21.9 $\pm$ 0.5 \cite{Pack2003}\\
 \hline
BaTiO$_3$ &$\mathrm{BaTiO_3}$            & uniaxial  & 4mm (room temp.)         & 0.40$\sim$9.00               & $d_{32(1.06)}$ = 14.4 $\pm$ 2.5\\
 \hline
MgBaF$_4$ &$\mathrm{MgBaF_4}$            & biaxial   & mm2         & 0.14$\sim$10.0               & $d_{32(1.064)}$ = 0.039 \\
 \hline
PMN-0.38PT \cite{He2006JAP} &  \tiny $\mathrm{0.62Pb(Mg_{1/3}Nb_{2/3})O_3-0.38PbTiO_3} $ & uniaxial  &4mm   & 0.3$\sim$11.0 &$^\dag$$ d_{33(1.064)}$ = 12.6 \\
\hline
OP-ZnSe \cite{Tropf1995OG} & $\mathrm{ZnSe}$               & isotropic & $\bar4$3m   & 0.45$\sim$18.0               & $d_{36(0.852)}$ = 53.8 \\
\hline \hline
\end{tabular}
\caption{\label{table:CrystalSummary} Main properties of the 10 uniaxial birefringent crystals (part I), 4 biaxial birefringent crystals (part II), and 8 periodic poling crystals (part III) discussed in this work, including the chemical formula, the axis (uniaxial, biaxial, or isotropic), the point group, the transparency range $\lambda_{transp.}$, and the maximal nonlinear coefficient $d_{max}$ at different wavelength (in $\mu$m in the bracket). Most of the data were obtained from Refs. \cite{Dmitriev1999, Nikogosyan2005}. The BaTiO$_3$ crystal has different types of point group at different temperatures. 
*$d_+$=$\left|d_{31}sin \theta \right|+\left|d_{22} cos \theta\right|$.
$^\dag$ The calculated result is from \cite{Kundys2020AQT} according to the method from \cite{Wang1999PRB}.
 }
\end{table*}

\section{Theory}\label{sec2}
\subsection{The characteristics of 22 kinds of nonlinear crystals}
We investigate 22 kinds of nonlinear crystals in this work. Table \ref{table:CrystalSummary} summarizes them from several perspectives: axial type, point group, transparency range, and the maximal nonlinear coefficient. We separate these crystals into three categories by their axial type or phase-matching form in order to illustrate the result more clearly in the next section. Most Sellmeier equations were obtained from Refs. \cite{Dmitriev1999, Nikogosyan2005}, and we have updated some Sellmeier equations with the latest references and concluded in Tab.\ref{table:CrystalSummary}.

In the first category, 9 birefringent uniaxial crystals are listed in the table. They have a very large transparency range up to 20 $\mu$m except CMTC and LiIO$_3$. We show  4 birefringent biaxial crystals in the second category, these four crystals can be written as Li$M$$X$ ($M$ = In, Ga and $X$ = S, Se). Here, In and Ga are elements of group \uppercase\expandafter{\romannumeral+3}A in the periodic table; S and Se are elements of group \uppercase\expandafter{\romannumeral+6}A. They are all mm2 point group. With their transparency range up to 14 $\mu$m, they have similar properties and can perform many applications in the MIR band. The other 8 periodic poling crystals  usually realize their phase-matching by the QPM method, so we discuss them in the section \Ref{sec3B} and \Ref{sec4} in detail.

\subsection{The GVM theory of spectrally-pure single-photon states generation}
The SPDC process generates the biphoton state $\vert\psi\rangle$, which can be written as
\begin{equation}\label{eq1}
\vert\psi\rangle=\int_0^\infty\int_0^\infty\,\mathrm{d}\omega_s\,\mathrm{d}\omega_if(\omega_s,\omega_i)\hat a_s^\dag(\omega_s)\hat a_i^\dag(\omega_i)\vert0\rangle\vert0\rangle,
\end{equation}
where $\hat a^\dag$ is the creation operator; $\omega$ is the angular frequency, and the subscripts $s$ and $i$ denote the signal and idler photon.
The joint spectral amplitude (JSA) $f(\omega_s,\omega_i)$ can be obtained by multiplying the pump envelope function (PEF) $\alpha(\omega_s,\omega_i)$ and the phase matching function (PMF) $\phi(\omega_s,\omega_i)$, i.e., $f(\omega_s,\omega_i)= \alpha(\omega_s,\omega_i) \times \phi(\omega_s,\omega_i).$

For PEF, it is usuaaly a Gaussian distribution and can be expressed as \cite{Mosley2008NJP},
\begin{equation}\label{eq20}
\alpha(\omega_s, \omega_i)=\exp[-\frac{1}{2}\left(\frac{\omega_s+\omega_i-\omega_{p0}}{\sigma_p}\right)^2],
\end{equation}
where $\sigma_p$ is the bandwidth of the pump;
$\omega_{p0}$ is the center frequency of the pump;
the full-width at half-maximum (FWHM) is FWHM$_\omega$=2$\sqrt{\ln(2)} \sigma_p \approx 1.67\sigma_p$.

If we use wavelengths as the variable by $\omega=2 \pi c/\lambda$ for ease of calculation, the PEF can be rewritten as
\begin{equation}\label{eq21}
\alpha(\lambda_s, \lambda_i)= \exp \left\{  -\frac{1}{2} \left( \frac{{1/\lambda _s  + 1/\lambda _i  - 1/(\lambda _0 /2)}}{{\Delta \lambda /[(\lambda _0 /2)^2  - (\Delta \lambda /2)^2 ]}}  \right) \right\},
\end{equation}
where  $\lambda _0 /2 $ is the central wavelength of the pump; $\Delta \lambda$ is the bandwidth of wavelength and $\sigma _p = \frac{{2\pi c}}{{\Delta \lambda /[(\lambda _0 )^2  - (\Delta \lambda /2)^2 ]}}$,
where $c$ is the light speed.

For $\Delta \lambda << \lambda _0$, The FWHM of the pump at intensity level is FWHM$_\lambda\approx 2\sqrt{\ln(2)} \Delta \lambda  \approx 1.67\Delta \lambda $.

By assuming a flat phase distribution, the PMF can be written as an \textit{sinc} function shape \cite{Mosley2008NJP},
\begin{equation}\label{eq22}
\phi(\omega_s,\omega_i)=\operatorname{sinc}\left(\frac{\Delta kL}{2}\right),
\end{equation}
where $L$ is the length of crystal, $\Delta k$ is the wave vector  mismatch. For QPM case, $\Delta k=k_p-k_i-k_s\pm \frac{2 \pi}{\Lambda}$, and $k=\frac{2 \pi n (\lambda)}{\lambda}$ is the wave vector. The refractive index $n(\lambda)$ is a function of wavelength $\lambda$. 
$\Lambda$ is the poling period, and $\Lambda = \frac{2\pi}{\lvert k_p-k_i-k_s \rvert}$.
For BPM case, $\Delta k=k_p-k_i-k_s$, and $k=\frac{2 \pi n (\lambda, \theta, \varphi)}{\lambda}$. For ordinary ray (o-ray), the refractive index $n_o (\lambda )$  is a function of wavelength $\lambda$. While for extraordinary ray (e-ray), the refractive index $n_e (\lambda, \theta, \varphi)$ is a function of polar angle $\theta$, azimuth angle $\varphi$ and wavelength $\lambda$.

According to the refractive index coordinate in Appendix of Ref. \cite{Jin2020QUE}, $\theta$ is the polar angle between the optical axis of the crystal and the light propagation direction, $\varphi$ is the azimuth angle in the $xy$ plane. For uniaxial crystals, $\theta$ is the cutting angle of the crystals. For biaxial crystals, when light propagates in the $xz$ plane, $\varphi$=0$^\circ$, $\theta$ is the cutting angle; when light propagates in the $yz$ plane, $\varphi$=90$^\circ$, $\theta$ is the cutting angle; when light propagates in the $xy$ plane, $\theta$=90$^\circ$, $\varphi$ is the cutting angle.

When the $\Delta k$=0, the phase-matching condition is satisfied. Under on this precondition, we consider the GVM condition to prepare an intrinsic spectrally-pure state.
The  angle $\theta_{PMF}$ between the positive direction of the horizontal axis and the ridge direction of the PEF is determined by \cite{Jin2013OE}:
\begin{equation}\label{eq4}
\tan(\theta_{PMF})=-\left( \frac{V_{g,p}^{-1}(\omega_p)-V_{g,s}^{-1}(\omega_s)}{V_{g,p}^{-1}(\omega_p)-V_{g,i}^{-1}(\omega_i)} \right),
\end{equation}
where $V_{g,\mu}=\frac{d\omega}{dk_\mu(\omega)}=\frac{1}{k_\mu^\prime(\omega)},(\mu=p, s, i)$ is the group velocity of the pump, the signal and the idler.

We consider three kinds of GVM conditions \cite{Jin2019PRApl}.
The GVM$_1$ condition ($\theta_{PMF}$ = 0$^\circ$) is
\begin{equation}\label{gvm1}
V_{g,p}^{-1}(\omega_p)=V_{g,s}^{-1}(\omega_s).
\end{equation}
The GVM$_2$ condition ($\theta_{PMF}$ = 90$^\circ$) is 
\begin{equation}\label{gvm2}
V_{g,p}^{-1}(\omega_p)=V_{g,i}^{-1}(\omega_i).
\end{equation}
The GVM$_3$ condition ($\theta_{PMF}$ = 45$^\circ$) is
\begin{equation}\label{gvm3}
2V_{g,p}^{-1}(\omega_p)=V_{g,s}^{-1}(\omega_s)+V_{g,i}^{-1}(\omega_i).
\end{equation}

The pure state not only can be prepared through these three GVM conditions but also all the conditions that the $\theta_{PMF}$ angles are between 0 and 90$^\circ$ \cite{Graffitti2018PRA,Quesada2018}.
Since these three GVM conditions are listed in Eqs. (\ref{gvm1}-\ref{gvm3}) are the most widely-used cases in the experiment, we mainly consider these three conditions within this work.
Besides, the degenerate or nondegenerate case of other $\theta_{PMF}$ under type-II and type-0 phase-matching conditions will be illustrated in section \ref{sec3C}.

\begin{table*}[tbp]
\centering
\begin{tabular}{ c|c c c}
\hline \hline
Name&GVM$_1$  (purity $ \approx $ 0.97)& GVM$_2$  (purity $ \approx $ 0.97) & GVM$_3$  (purity $ \approx $ 0.82) \\
\hline
CMTC*& \tabincell{l}{ $\lambda_p$ = 649 nm, $\lambda_{s,i}$ = 1298 nm \\  $\theta$ = 41.2$^\circ$,  $d_\textrm{eff}$ = $-$4.46 pm/V}
     & \tabincell{l}{ $\lambda_p$ = 1156 nm, $\lambda_{s,i}$ = 2312 nm \\  $\theta$ = 43.6$^\circ$,  $d_\textrm{eff}$ = $-$3.63 pm/V}
      & \tabincell{l}{$\lambda_p$ = 829 nm, $\lambda_{s,i}$ = 1658 nm \\  $\theta$ = 38.2$^\circ$,  $d_\textrm{eff}$ = $-$4.13 pm/V}  \\
\hline
THI$^\dag$& \tabincell{l}{$\lambda_p$ = 2841 nm, $\lambda_{s,i}$ = 5682 nm \\  $\theta$ = 29.3$^\circ$,  $d_\textrm{eff}$ = unknown }
      &  \tabincell{l}{$\lambda_p$ = 4822 nm, $\lambda_{s,i}$ = 9644 nm \\  $\theta$ = 29.3$^\circ$,  $d_\textrm{eff}$ = unknown }
      &  \tabincell{l}{$\lambda_p$ = 3705 nm, $\lambda_{s,i}$ = 7410 nm \\  $\theta$ = 27.2$^\circ$,  $d_\textrm{eff}$ = unknown }  \\
\hline
LiIO${_3}$& \tabincell{l}{$\lambda_p$ = 835 nm, $\lambda_{s,i}$ = 1670 nm \\  $\theta$ = 29.3$^\circ$,  $d_\textrm{eff}$ = 0.09 pm/V}
      &  \tabincell{l}{$\lambda_p$ = 1460 nm, $\lambda_{s,i}$ = 2920 nm \\  $\theta$ = 29.7$^\circ$,  $d_\textrm{eff}$ = 0.09 pm/V}
      &  \tabincell{l}{$\lambda_p$ = 1088 nm, $\lambda_{s,i}$ = 2176 nm \\  $\theta$ = 27.2$^\circ$,  $d_\textrm{eff}$ = 0.09 pm/V}  \\
\hline
AAS& \tabincell{l}{$\lambda_p$ = 2151 nm, $\lambda_{s,i}$ = 4302 nm \\  $\theta$ = 22.3$^\circ$,  $d_\textrm{eff}$ = 0.16 pm/V}
        &  \tabincell{l}{$\lambda_p$ = 3617 nm, $\lambda_{s,i}$ = 7234 nm \\  $\theta$ = 22.3$^\circ$,  $d_\textrm{eff}$ = 0.15 pm/V}
        &  \tabincell{l}{$\lambda_p$ = 2793 nm, $\lambda_{s,i}$ = 5586 nm \\  $\theta$ = 20.7$^\circ$,  $d_\textrm{eff}$ = 0.16 pm/V}  \\
\hline
HGS & \tabincell{l}{$\lambda_p$ = 1704 nm, $\lambda_{s,i}$ = 3408 nm \\  $\theta$ = 60.1$^\circ$,  $d_\textrm{eff}$ = 0.29 pm/V}
      &  \tabincell{l}{$\lambda_p$ = 2819 nm, $\lambda_{s,i}$ = 5638 nm \\  $\theta$ = 59.8$^\circ$,  $d_\textrm{eff}$ = 0.29 pm/V}
      &  \tabincell{l}{$\lambda_p$ = 2206 nm, $\lambda_{s,i}$ = 4412 nm \\  $\theta$ = 54.2$^\circ$,  $d_\textrm{eff}$ = 0.32 pm/V}  \\
\hline
CGA& \tabincell{l}{$\lambda_p$ = 3692 nm, $\lambda_{s,i}$ = 7384 nm \\  $\theta$ = 54.6$^\circ$,  $d_\textrm{eff}$ = 0.02 pm/V}
      &  \tabincell{l}{$\lambda_p$ = 5825 nm, $\lambda_{s,i}$ = 11650 nm \\  $\theta$ = 53.9$^\circ$,  $d_\textrm{eff}$ = 0.02 pm/V}
      &  \tabincell{l}{$\lambda_p$ = 4690 nm, $\lambda_{s,i}$ = 9380nm \\  $\theta$ = 49.8$^\circ$,  $d_\textrm{eff}$ = 0.02 pm/V}  \\
\hline
TAS& \tabincell{l}{$\lambda_p$ = 3620 nm, $\lambda_{s,i}$ = 7240nm \\  $\theta$ = 27.5$^\circ$,  $d_\textrm{eff}$ = 0.24 pm/V}
      &  \tabincell{l}{$\lambda_p$ = 5535 nm, $\lambda_{s,i}$ = 11070 nm \\  $\theta$ = 27.0$^\circ$,  $d_\textrm{eff}$ = 0.23 pm/V}
      &  \tabincell{l}{$\lambda_p$ = 4570 nm, $\lambda_{s,i}$ = 9140 nm \\  $\theta$ = 25.6$^\circ$,  $d_\textrm{eff}$ = 0.24 pm/V}  \\
\hline
AGS& \tabincell{l}{$\lambda_p$ = 1688 nm, $\lambda_{s,i}$ = 3376nm \\  $\theta$ = 53.7$^\circ$,  $d_\textrm{eff}$ = 0.14 pm/V}
      &  \tabincell{l}{$\lambda_p$ = 2845 nm, $\lambda_{s,i}$ = 5690 nm \\  $\theta$ = 53.9$^\circ$,  $d_\textrm{eff}$ = 0.14 pm/V}
      &  \tabincell{l}{$\lambda_p$ = 2187 nm, $\lambda_{s,i}$ = 4374 nm \\  $\theta$ = 48.9$^\circ$,  $d_\textrm{eff}$ = 0.15 pm/V}  \\
\hline
AGSe& \tabincell{l}{$\lambda_p$ = 2457 nm, $\lambda_{s,i}$ = 4914nm \\  $\theta$ = 79.9$^\circ$,  $d_\textrm{eff}$ = 0.11 pm/V}
      &  \tabincell{l}{$\lambda_p$ = 4079 nm, $\lambda_{s,i}$ = 8158 nm \\  $\theta$ = 81.9$^\circ$,  $d_\textrm{eff}$ = 0.13 pm/V}
      &  \tabincell{l}{$\lambda_p$ = 3136 nm, $\lambda_{s,i}$ = 6272 nm \\  $\theta$ = 67.2$^\circ$,  $d_\textrm{eff}$ = 0.25 pm/V}  \\
\hline
GaSe& \tabincell{l}{$\lambda_p$ = 2189 nm, $\lambda_{s,i}$ = 4378 nm \\  $\theta$ = 16.1$^\circ$,  $d_\textrm{eff}$ = 0.51 pm/V}
      &  \tabincell{l}{$\lambda_p$ = 3657 nm, $\lambda_{s,i}$ = 7314 nm \\  $\theta$ = 16.1$^\circ$,  $d_\textrm{eff}$ = 0.49 pm/V}
      &  \tabincell{l}{$\lambda_p$ = 2833 nm, $\lambda_{s,i}$ = 5666 nm \\  $\theta$ = 15.0$^\circ$,  $d_\textrm{eff}$ = 0.51 pm/V}  \\
\hline \hline
\end{tabular}
\caption{\label{table:uniaxial}
Three kinds of GVM conditions for 10 uniaxial BPM crystals. $\lambda_{p(s,i)}$ is the GVM wavelength for the pump (signal, idler). $\theta$ is the phase-matching angle and $d_\textrm{eff}$ is the effective nonlinear coefficient. The Sellmeier equations are obtained from Refs. \cite{Dmitriev1999, Nikogosyan2005}. 
Most of the $d_\textrm{eff}$ values can be obtained from the SNLO $v78$ software package, developed by AS-Photonics, LLC \cite{SNLO78}.
*The $d_\textrm{eff}$ value for CMTC is not available from SNLO, we have calculated the $d_\textrm{eff}$ using the method in the Appendix of Ref. \cite{Jin2020QUE} and considering Miller's rule \cite{smith2018crystal}. 
$^\dag$ The $d_\textrm{eff}$ value for THI is unknown.
}
\end{table*}

\begin{table*}[!tbp]
\centering
\begin{tabular}{ c|c c c}
\hline \hline
Name&GVM$_1$  (purity $ \approx $ 0.97)& GVM$_2$  (purity $ \approx $ 0.97) & GVM$_3$  (purity $ \approx $ 0.82) \\
\hline
\hline
LIS$-{xy}$& \tabincell{l}{$\lambda_p$ = 1457 nm, $\lambda_{s,i}$ = 2914 nm \\  $\varphi$ = 56.7$^\circ$,  $d_\textrm{eff}$ = 6.01 pm/V}
     & \tabincell{l}{$\lambda_p$ = 2473 nm, $\lambda_{s,i}$ = 4946 nm \\  $\varphi$ = 56.7$^\circ$,  $d_\textrm{eff}$ = 5.74 pm/V}
      & \tabincell{l}{$\lambda_p$ = 1901 nm, $\lambda_{s,i}$ = 3802 nm \\  $\varphi$ = 49.5$^\circ$,  $d_\textrm{eff}$ = 6.05 pm/V}
        \\
\hline
LISe$-{xy}$ &\tabincell{l}{$\lambda_p$ = 1912 nm, $\lambda_{s,i}$ = 3824 nm \\  $\varphi$ = 45.8$^\circ$,  $d_\textrm{eff}$ = 9.65 pm/V}
     &\tabincell{l}{$\lambda_p$ = 3205 nm, $\lambda_{s,i}$ = 6410 nm \\  $\varphi$ = 45.6$^\circ$,  $d_\textrm{eff}$ = 9.31 pm/V}
      &\tabincell{l}{$\lambda_p$ = 2491 nm, $\lambda_{s,i}$ = 4982 nm \\  $\varphi$ = 40.7$^\circ$,  $d_\textrm{eff}$ = 9.19 pm/V}
      \\
\hline
LGS$-{xy}$& \tabincell{l}{$\lambda_p$ = 1347 nm, $\lambda_{s,i}$ = 2696 nm \\  $\varphi$ = 55.5$^\circ$,  $d_\textrm{eff}$ = 5.24 pm/V}
     & \tabincell{l}{$\lambda_p$ = 2282 nm, $\lambda_{s,i}$ = 4564 nm \\  $\varphi$ = 55.1$^\circ$,  $d_\textrm{eff}$ = 5.02 pm/V}
      & \tabincell{l}{$\lambda_p$ = 1767 nm, $\lambda_{s,i}$ = 3534 nm \\  $\varphi$ = 49.7$^\circ$,  $d_\textrm{eff}$ = 5.19 pm/V}  \\
\hline
LGSe$-{xy}$& \tabincell{l}{$\lambda_p$ = 1641 nm, $\lambda_{s,i}$ = 3282 nm \\  $\varphi$ = 47.8$^\circ$,  $d_\textrm{eff}$ = 8.37 pm/V}
      & \tabincell{l}{$\lambda_p$ = 2729 nm, $\lambda_{s,i}$ = 5258 nm \\  $\varphi$ = 47.5$^\circ$,  $d_\textrm{eff}$ = 8.06 pm/V}
       & \tabincell{l}{$\lambda_p$ = 2129 nm, $\lambda_{s,i}$ = 4258 nm \\  $\varphi$ = 43.3$^\circ$,  $d_\textrm{eff}$ = 8.38 pm/V}  \\
\hline
\hline
\end{tabular}
\caption{\label{table:biaxial} Three kinds of GVM conditions for 4 biaxial BPM crystals with light propagating in the $xy$ plane.
$\varphi$ is the azimuth angle.
The $d_\textrm{eff}$ are obtained from SNLO $v78$ \cite{SNLO78}.
}
\end{table*}

\begin{figure}[tbp]
\centering
\includegraphics[width=8.5cm]{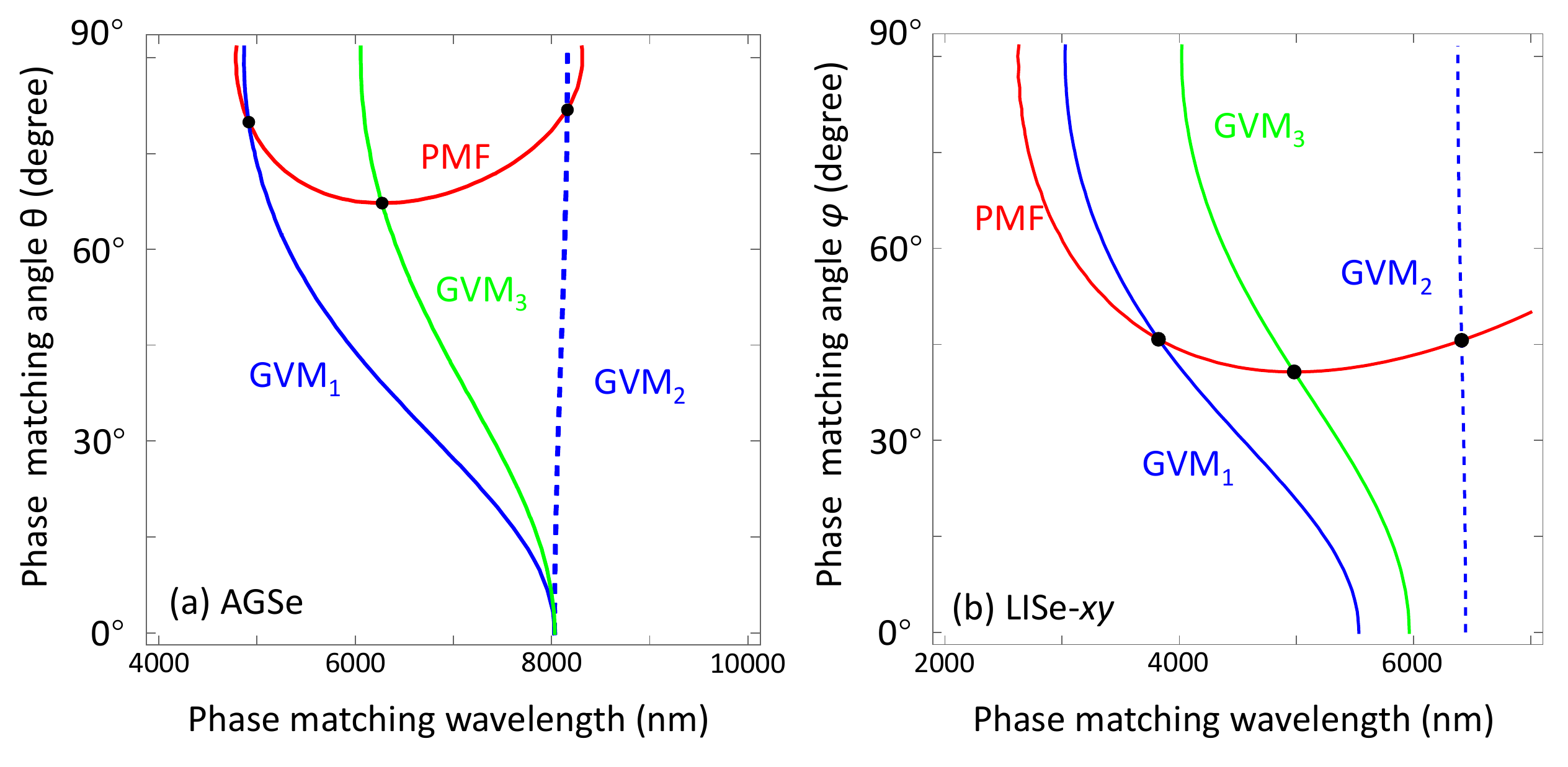}
\caption{The phase-matching function and group-velocity matching functions (GVM$_{1(2,3)}$) for different signal/idler wavelength and phase-matching angle (polar angle) $\theta$ for AGSe crystal (a) and phase-matching (azimuth angle) angle $\varphi$ for LISe crystal in the xy plane (b). In this calculation, we consider the Type-II phase-matching condition with collinear and wavelength-degenerate (2 $\lambda_p$ = $\lambda_s$ = $\lambda_i$ ) configuration.}\label{GVM}
\end{figure}

\begin{figure}[!tbp]
\centering
\includegraphics[width=8.5cm]{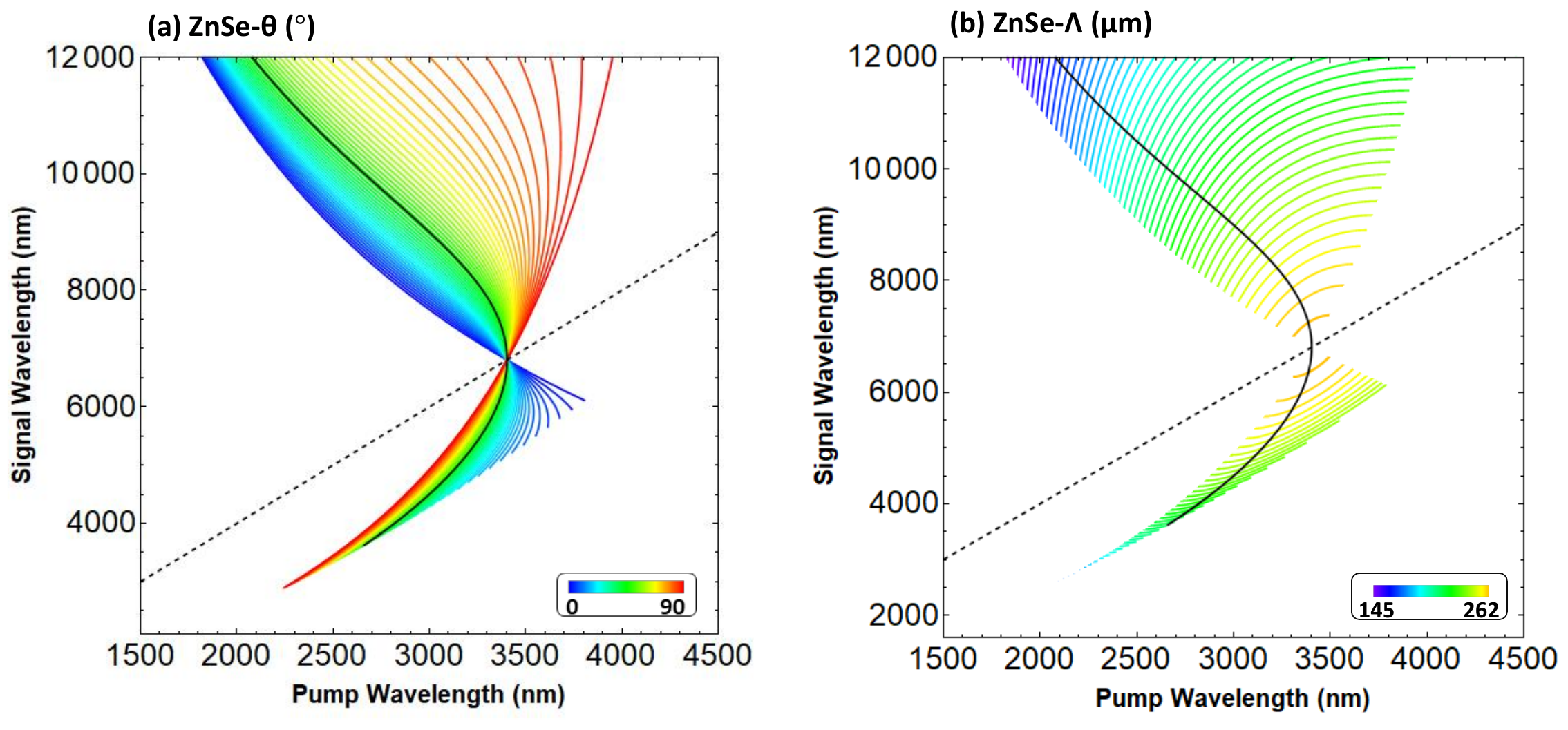}
\caption{
(a, b): The GVM angle $\theta_{PMF}$ and the corresponding poling period $\Lambda$ for QPM crystal OP-ZnSe. The solid black line indicates $\theta_{PMF}$ = 45$^\circ$, and the dashed black line indicates the degenerate case 2$\lambda_p$=$\lambda_s$. 
In this figure, 
the QPM nondegenerate wavelengths case is under type-0 phase-matching condition.}\label{nondeg}
\end{figure}

\begin{table*}[!htbp]
\centering
\begin{tabular}{ c|c c c}
\hline \hline
Name&GVM$_1$  (purity $ \approx $ 0.97)& GVM$_2$  (purity $ \approx $ 0.97) & GVM$_3$  (purity $ \approx $ 0.82) \\
\hline
LT& \tabincell{l}{ $\lambda_p$ = 1279 nm, $\lambda_{s,i}$ = 2558 nm \\  $\Lambda$ = 33.7 $\mu$m,  $d_\textrm{eff}$= 0.27 pm/V}
     & \tabincell{l}{ $\lambda_p$ = 1320 nm, $\lambda_{s,i}$ = 2640 nm \\  $\Lambda$ = 33.7 $\mu$m,  $d_\textrm{eff}$ = 0.27 pm/V}
      & \tabincell{l}{$\lambda_p$ = 1299 nm, $\lambda_{s,i}$ = 2598 nm \\  $\Lambda$ = 33.7 $\mu$m,  $d_\textrm{eff}$ = 0.27 pm/V}  \\
\hline
LN& \tabincell{l}{$\lambda_p$ = 1341 nm, $\lambda_{s,i}$ = 2682 nm \\  $\Lambda$ = 14.7 $\mu$m,  $d_\textrm{eff}$ = 2.70 pm/V}
      &  \tabincell{l}{$\lambda_p$ = 2015 nm, $\lambda_{s,i}$ = 4030 nm \\  $\Lambda$ = 15.2 $\mu$m,  $d_\textrm{eff}$ = 2.50 pm/V}
      &  \tabincell{l}{$\lambda_p$ = 1709 nm, $\lambda_{s,i}$ = 3418 nm \\  $\Lambda$ = 15.5 $\mu$m,  $d_\textrm{eff}$ = 2.60 pm/V}  \\
\hline
KTP& \tabincell{l}{$\lambda_p$ = 613 nm, $\lambda_{s,i}$ = 1224 nm \\  $\Lambda$ = 70.2 $\mu$m,  $d_\textrm{eff}$ = 2.50 pm/V}
        &  \tabincell{l}{$\lambda_p$ = 1169 nm, $\lambda_{s,i}$ = 2338 nm \\  $\Lambda$ = 72.2 $\mu$m,  $d_\textrm{eff}$ = 2.40 pm/V}
        &  \tabincell{l}{$\lambda_p$ = 792 nm, $\lambda_{s,i}$ = 1584 nm \\  $\Lambda$ = 45.0 $\mu$m,  $d_\textrm{eff}$ = 2.40 pm/V}  \\
\hline
KN & \tabincell{l}{$\lambda_p$ = 1412 nm, $\lambda_{s,i}$ = 2824 nm \\  $\Lambda$ = 6.2 $\mu$m,  $d_\textrm{eff}$ = 5.20 pm/V}
      &  \tabincell{l}{$\lambda_p$ = 1869 nm, $\lambda_{s,i}$ = 3738 nm \\  $\Lambda$ = 6.1 $\mu$m,  $d_\textrm{eff}$ = 5.00 pm/V}
      &  \tabincell{l}{$\lambda_p$ = 1605 nm, $\lambda_{s,i}$ = 3210 nm \\ $\Lambda$ = 6.3 $\mu$m,  $d_\textrm{eff}$ = 5.10 pm/V}  \\
\hline
BaTiO$_3$*& \tabincell{l}{$\lambda_p$ = 1518 nm, $\lambda_{s,i}$ = 3036 nm \\  $\Lambda$ = 23.3 $\mu$m,  $d_\textrm{eff}$ = 10.13 pm/V}
      &  \tabincell{l}{$\lambda_p$ = 1993 nm, $\lambda_{s,i}$ = 3986 nm \\  $\Lambda$ = 23.3 $\mu$m,  $d_\textrm{eff}$ = 9.17 pm/V}
      &  \tabincell{l}{$\lambda_p$=1740 nm, $\lambda_{s,i}$=3480 nm \\ $\Lambda$ = 23.8 $\mu$m,  $d_\textrm{eff}$ = 9.69 pm/V}  \\
\hline
MgBaF$_4$*&  \tabincell{l}{$\lambda_p$ = 989 nm, $\lambda_{s,i}$ = 1978 nm \\  $\Lambda$ = 948.5 $\mu$m,  $d_\textrm{eff}$ = 0.04 pm/V}
      &  not satisfied
      &  \tabincell{l}{$\lambda_p$ = 1390 nm, $\lambda_{s,i}$ = 2780 nm \\  $\Lambda$ = 680.6 $\mu$m,  $d_\textrm{eff}$ = 0.04 pm/V}  \\
\hline
PMN-0.38PT$^\dag$&  \tabincell{l}{$\lambda_p$ = 2810 nm, $\lambda_{s,i}$ = 5620 nm \\  $\Lambda$ = 1301.38 $\mu$m,  $d_\textrm{eff}$ = unknown}
     &  not satisfied
      &  \tabincell{l}{$\lambda_p$ = 3972 nm, $\lambda_{s,i}$ = 7944 nm \\  $\Lambda$ = 917.83 $\mu$m,  $d_\textrm{eff}$ = unknown}  \\
\hline
OP-ZnSe&
      &  \tabincell{l}{$\lambda_p$ = 3403 nm, $\lambda_{s,i}$ = 6806 nm \\  $\Lambda$ = 262.97 $\mu$m,  $d_\textrm{eff}$ = 19.1 pm/V}
      &   \\
\hline \hline
\end{tabular}
\caption{\label{table:QPMaxial}
Three kinds of GVM conditions for 8 QPM crystals. $\lambda_{p(s,i)}$ is the GVM wavelength for the pump (signal, idler). $\Lambda$ is the poling period and $d_\textrm{eff}$ is the effective nonlinear coefficient. The Sellmeier equations are obtained from Refs. \cite{Dmitriev1999, Nikogosyan2005}.
Most of the $d_\textrm{eff}$ values can be obtained from the SNLO $v78$ software package, developed by AS-Photonics, LLC \cite{SNLO78}.
*The $d_\textrm{eff}$ values for BaTiO$_3$ and MgBaF$_4$ are not available from SNLO, so we have calculated them using the method in the Appendix of Ref. \cite{Jin2020QUE} and considering Miller's rule \cite{smith2018crystal}. $^\dag$ The $d_\textrm{eff}$ value for PMN-0.38PT is unknown.
}
\end{table*}

\begin{figure*}[!htbp]
\centering
\includegraphics[width=16cm]{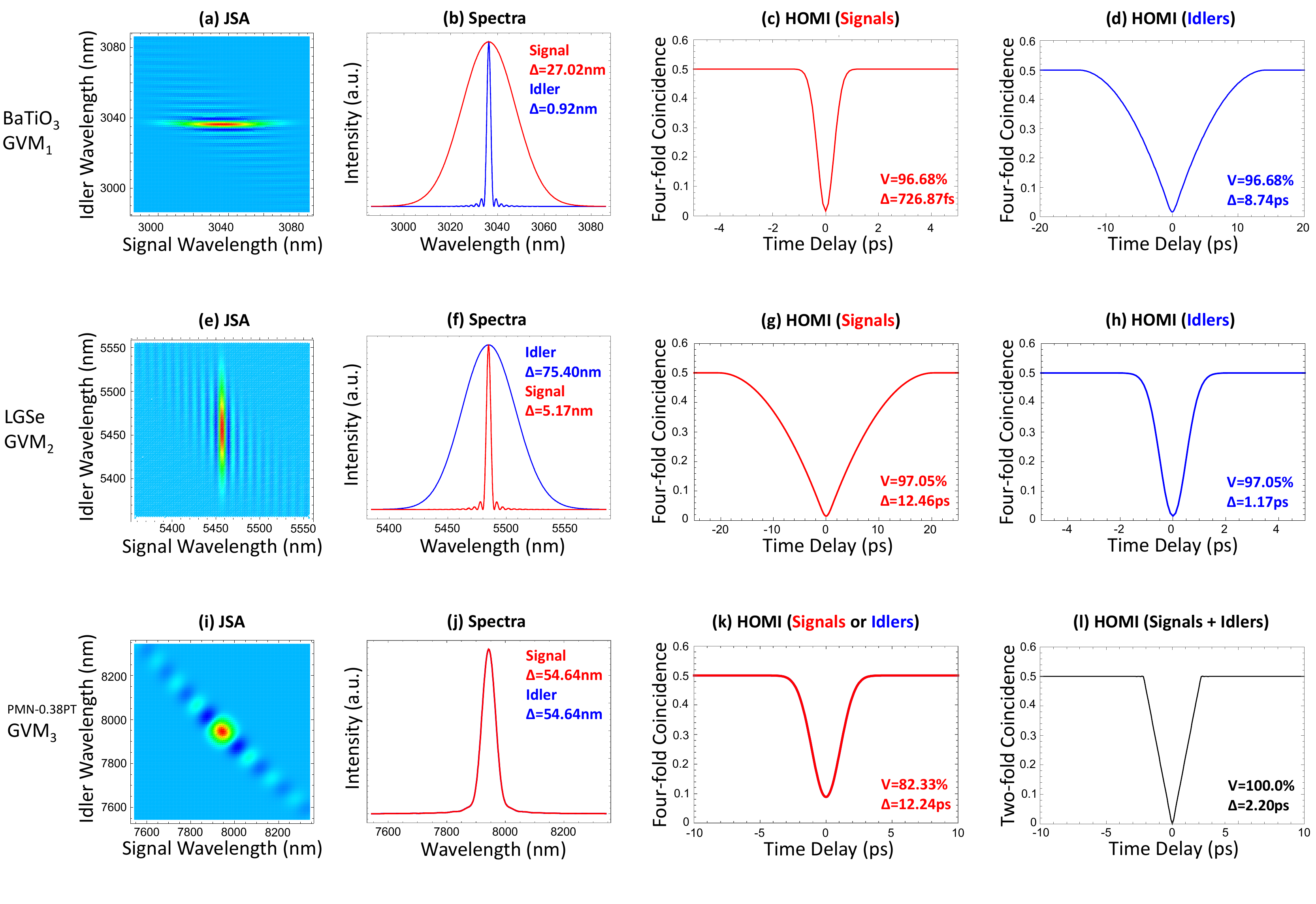}
\caption{
(a, e, i) are the JSAs; (b, f, j) are the spectra; and (c, d, g, h, k, l) are the HOM interference patterns. The FWHM of the spectra ($\Delta$) for the signal and the idler, the visibility (V) and the FWHM ($\Delta$) of two-fold and four-fold HOM interference are shown in the figures. The parameters of $L=100$ mm and $\Delta \lambda=4$ nm (FWHM = 6.66 nm), $L=200$ mm and $\Delta \lambda=8$ nm (FWHM = 13.32 nm), $L=100$ mm and $\Delta \lambda=11$ nm (FWHM = 18.32 nm) are adopted for BaTiO$_3$, LGSe, and PMN-0.38PT, respecitvely. Note that for all the calculations of purity, we use a grid size of 200$\times$200 for all the JSAs.
 } \label{HOM}
\end{figure*}

\begin{figure*}[tbp]
\centering
\includegraphics[width=15cm]{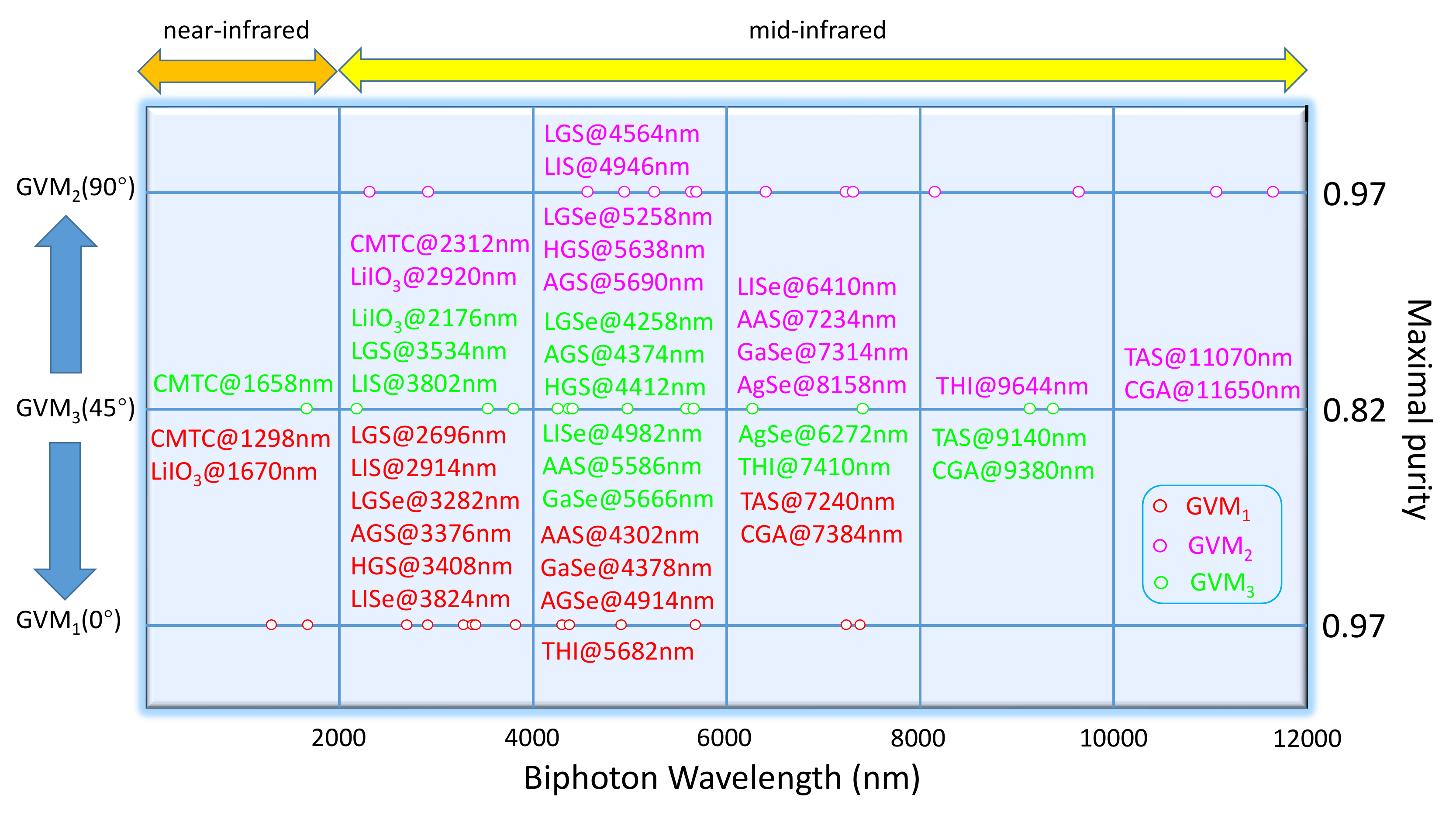}
\caption{The result of all the BPM crystals for three kinds of GVM conditions. GVM$_{1(2)}$ can achieve a purity of 0.97. GVM$_3$ can achieve a purity of 0.82. The GVM angle increases from 0$^\circ$ to 45$^\circ$ and ends up at 90$^\circ$ on the left axis for GVM$_{1(2,3)}$ condition. The GVM$_1$ condition (red points) can cover the wavelength range from 1298 nm to 7384 nm. The GVM$_2$ condition (magenta points) can cover the wavelength range from 2312 nm to 11650 nm. The GVM$_3$ condition (green points) can cover the wavelength range from 1658 nm to 9380 nm.}\label{BPM-result}
\end{figure*}

\section{Calculation and Simulation}\label{sec3}
\subsection{Birefringent crystals}\label{sec3A}

Firstly, we consider the birefringent crystals with the BPM method. We assume the wavelength is degenerated, i.e., 2$\lambda_p$ = $\lambda_s$ = $\lambda_i$. For uniaxial crystals, negative uniaxial crystals satisfy the type-II SPDC with e$\rightarrow$o+e phase-matching interaction. Here  the pump and idler are extraordinary (e) beams, while the signal is ordinary (o) beam. In contrast, the positive uniaxial crystals can meet the Type-II SPDC with o$\rightarrow$o+e phase-matching interaction. Note all the simulations are based on collinear figuration.

10 kinds of uniaxial crystals are investigated in this work.
Taking AgSe crystal as an example, we plot the PMF and GVM$_{1(2,3)}$ conditions for different wavelengths and phase-matched angles in Fig. \ref{GVM} (a). The PMF (red) crosses the GVM$_{1(2,3)}$ (blue and green) at three black points, which meet the PMF and three kinds of GVM conditions simultaneously. The three points  associated with wavelengths  of 4914, 8158, and 6272 nm, respectively, and angles of 79.9$^\circ$, 81.9$^\circ$, and 67.2$^\circ$. Further, we calculate the other uniaxial crystals with the same method, and then we summarize the result in Table \ref{table:uniaxial}.
In Table \ref{table:uniaxial}, the down-converted photons have a wavelength range from 1298 to 11650 nm, which is in the near-infrared (NIR) and MIR bands. The corresponding spectral purity at GVM$_{1(2,3)}$ wavelengths is 0.97, 0.97, and 0.82, respectively.

For biaxial crystals, all the crystals we investigated can only satisfy the GVM conditions in the xy plane, with polar angle $\theta$ = 90$^\circ$. We study 4 kinds of biaxial crystals and choose the LISe crystal as an example, which represents the mm2 point group. The assignment of dielectric and crystallographic axes are X, Y, Z $\Rightarrow$ b, a, c. As shown in Fig. \ref{GVM} (b), in the xy plane the cross points reflect the GVM condition can be fulfilled at the wavelength of 3824, 6410, and 4982 nm, respectively.

All the biaxial crystals can prepare pure-state in the range from 2696 to 6410 nm in Tab. \ref{table:biaxial}. We can notice that the wavelength range is from 1298 to 7384 nm for the GVM$_{1}$ condition, from 2312 to 11650 nm for the GVM$_{2}$ condition, and from 1658 to 9380 nm for the GVM$_{3}$  condition (also shown in Fig. \ref{BPM-result}), which can meet the different application demand in NIR, MIR, and telecom wavelengths.

\begin{figure*}[tbp]
\centering
\includegraphics[width=16cm]{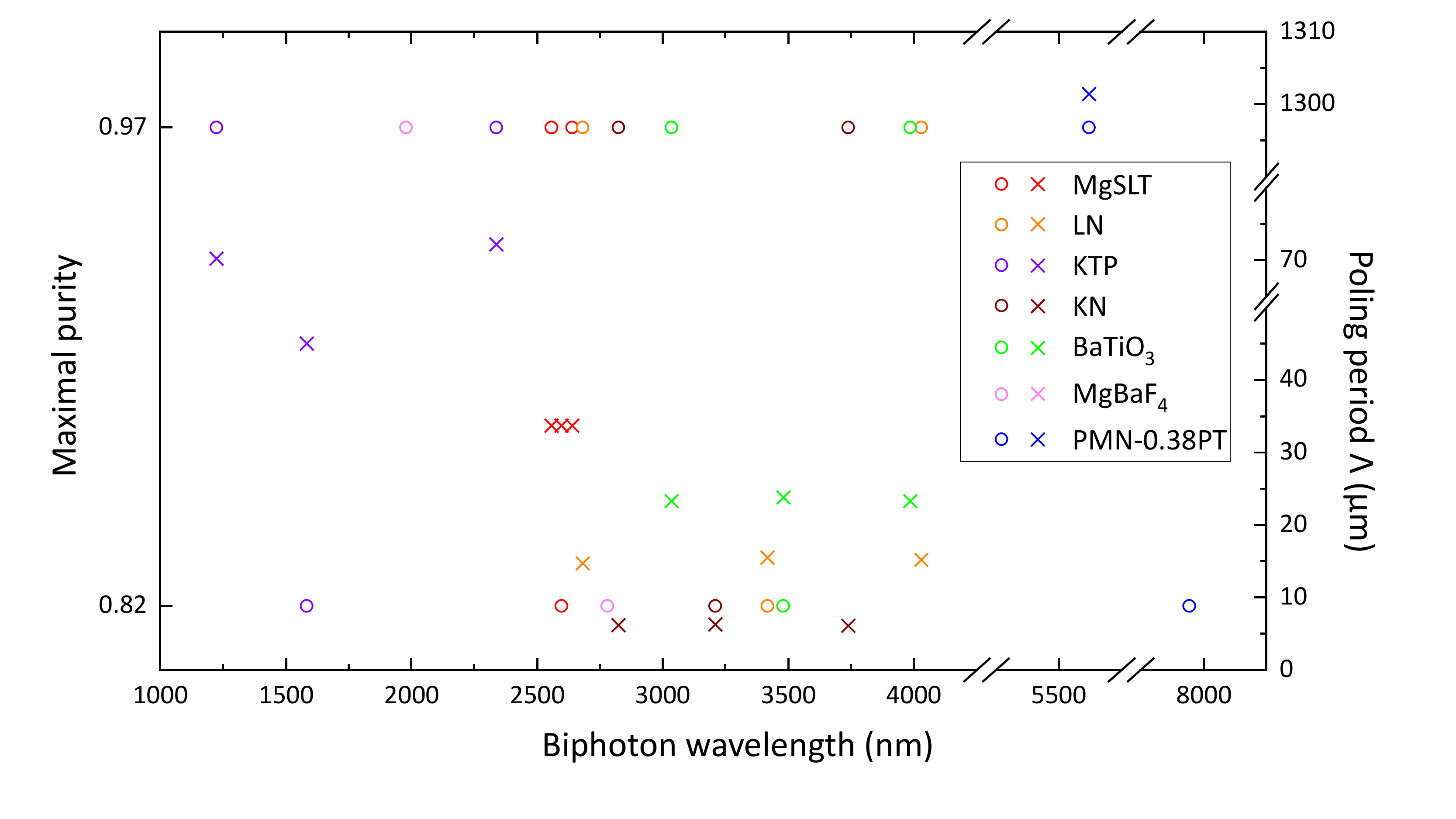}
\caption{The result of all the QPM crystals for three kinds of GVM conditions. GVM$_{1(2)}$ can achieve a purity of 0.97. GVM$_3$ can achieve a purity of 0.82. The wavelength versus achievable maximal spectral purity and poling period $\Lambda$ can be reflected on the scale of the left and right Y-axis.
The wavelength range is from 1224 nm to 7944 nm, and the poling period $\Lambda$ is from 6.1 $\mu$m to 1301.38 $\mu$m}\label{QPM-result}
\end{figure*}

\subsection{Periodic poling crystals}\label{sec3B}
In this section, we consider 8 periodic poling crystals with the QPM method, which has several advantages. For example, the largest component of the nonlinear coefficient matrix (usually $d_{33}$) can be utilized;  there is no walk-off angle so as to achieve good spatial mode;   it allows phase-matching interaction in isotropic media, in which the BPM is not applicable \cite{Hum2007}. The GVM wavelengths, the poling period $\Lambda$, and the effective nonlinear coefficient d$_{eff}$ are calculated and listed in Tab. \ref{table:QPMaxial}.
The LT, LN, KTP, and KN crystals are traditionally often-used QPM crystals \cite{Shimizu2009OE, Wang2020PPLN3D, Jin2016PRAppl, Lee2016AO}. Here, we find  the GVM$_2$ wavelengths are all above 2 $\mu$m.

The BaTiO$_3$ crystal shows a low birefringence, thus, only suitable for QPM method. With its high transmission in the IR range, it is possible to prepare pure-state at 3036, 3986, and 3480 nm, respectively.
The MgBaF$_4$ crystal can meet the GVM$_1$ condition at 1978 nm, and the GVM$_3$ condition at 2780 nm. Note that this crystal does not satisfy the GVM$_2$ condition.
The PMN-0.38PT is a functional ferroelectric material. The GVM condition only can be fulfilled at two wavelengths, i.e., 5620 nm and 7944 nm for GVM$_1$ and GVM$_3$ conditions.
The orientation-patterned zinc selenide (OP-ZnSe) is an isotropic semiconductor material, therefore, the QPM rather than BPM is applicable. OP-ZnSe has extremely high nonlinear coefficients. Since the crystal possesses only one refractive index, it can only perform type-0 SPDC, i.e., e$\rightarrow$e+e interaction, which will be discussed in the next section.
All the QPM crystals can prepare pure-state at the range from 1224 to 7944 nm, as listed in Tab. \ref{table:QPMaxial}.

\subsection{Wavelength nondegenerate case}\label{sec3C}

In this section, we focus on the wavelength nondegenerate case using QPM method.
We take OP-ZnSe as an example and calculate the  $\theta_{PMF}$ in the range of 0 and 90 degrees and the corresponding poling period $\Lambda$ in Fig. \ref{nondeg}.
The dashed black line  in Fig. \ref{nondeg} indicates the degenerate case, i.e., 2$\lambda_p$ = $\lambda_s$. For one fixed pump wavelength, we can find  $\theta_{PMF}$ and  $\Lambda$ of different signal wavelengths.

The OP-ZnSe crystal can only perform type-0 SPDC, i.e., e$\rightarrow$e+e interaction. Under the degenerate condition, the signal and the idler have the same group velocity, so pure state can not be prepared. The lines of all the angles $\theta_{PMF}$ converge in one point. At this point, all the GVM conditions are satisfied synchronously. Due to the singularity caused by these GVM conditions, the degenerate case at this point does not provide high purity, while the pure state can be prepared in the other area of different  $\theta_{PMF}$.

\subsection{HOM interference simulation}
The quality of the spectrally uncorrelated biphoton state can be tested by Hong-Ou-Mandel (HOM) interference. There are two kinds of HOM interference, the first one is the HOM interferences using signal and idler photons from the same SPDC source, with a typical setup shown in \cite{Hong1987}. In this case, the two-fold coincidence probability $ P_2(\tau )$ as a function of the time delay $\tau$ is given by \cite{Grice1997, Ou2007, Jin2015OE}:
\begin{equation}\label{eq:P2}
\begin{split}
P_2(\tau ) =  & \frac{1}{4} \int\limits_0^\infty  \int\limits_0^\infty  d\omega _s  d\omega _i \\
              & \left| {[f(\omega _s ,\omega _i ) - f(\omega _i ,\omega _s )e^{ - i(\omega _s  - \omega _i )\tau } ]} \right|^2.
\end{split}
\end{equation}
The second one is the HOM interference with two independent heralded single-photon sources, with a typical experimental setup shown in Refs. \cite{Mosley2008PRL, Jin2013PRA}.
In this interference, two signals $s_{1}$ and $s_{2}$ are sent to a beamsplitter for interference, and two idlers $i_{1}$ and $i_{2}$ are detected by single-photon detectors for heralding the signals.
The four-fold coincidence counts $P_4$ as a function of $\tau$ can be described by \cite{Ou2007, Jin2015OE}
\begin{equation}\label{eq:P4}
\begin{split}
P_4 (\tau )  = & \frac{1}{4}  \int_0^\infty \int_0^\infty \int_0^\infty \int_0^\infty d\omega _{s_1} d\omega _{s_2} d\omega _{i_1} d\omega _{i_2}  \\ & {\rm{|}}f_1 (\omega _{s_1} ,\omega _{i_1} )f_2 (\omega _{s_2} ,\omega _{i_2} )- \\ & f_1 (\omega _{s_2} ,\omega _{i_1} )f_2 (\omega _{s_1} ,\omega _{i_2} )e^{ - i(\omega _{s_2}  - \omega _{s_1} )\tau } {\rm{|}}^{\rm{2}},
\end{split}
\end{equation}
where $f_1$ and $f_2$ are the JSAs from the first and the second crystals.

We choose BiTaO$_3$, LGSe, and PMN-0.38PT as  examples to test the HOM interference.
Figure \ref{HOM} (a) shows JSA figure is generated from BiTaO$_3$ crystal, which is under the GVM$_1$ condition. BiTaO$_3$ crystal is a uniaxial QPM crystal. The JSA is obtained by using a pump laser with a bandwidth of $\Delta\lambda$ = 4 nm, and a crystal length L of 100 mm. The JSA  has a long stripe shape along the horizontal axis. As for the spectral distributions of the signal and the idler photons, we can obtain them by projecting the joint spectral intensity onto the horizontal and vertical axes. The FWHM of the signal (idler) is 27.02 nm (0.92 nm). Figure \ref{HOM} (c) shows the HOM pattern of two signals heralded by two idlers, the FWHM is 726.87 fs with visibility of 96.68\%. Figure \ref{HOM} (d) shows the HOM pattern of two heralded idlers with an FWHM of 8.74 ps and a visibility of 96.68\%.

For the GVM$_2$ condition, the result is on the second row of Fig. \ref{HOM}. We investigate a biaxial BPM crystal LGSe. The JSA shape is also a long stripe, but it locates along the vertical axis. The pump bandwidth $\Delta\lambda$ and the crystal length L of Fig.\ref{HOM} (e) are 8 nm and 200 mm. The FWHM of the signal (idler) is 5.17 nm (75.40 nm) for Fig.\ref{HOM} (f). The FWHM of the HOM pattern by two heralded signals (idler) is 12.46 ps (1.17 ps), and the visibility is 97.05\%, as shown in Fig.\ref{HOM} (g)(h).

For the GVM$_3$ condition, the result is on the third row of Fig. \ref{HOM}. We concentrate on PMN-0.38PT crystal. This crystal has been studied before, however, it only focuses on the GVM$_1$ and GVM$_2$ conditions \cite{Kundys2020AQT}. Here  we make a thorough study of the GVM$_3$ condition. In this case, the JSA shape is near-round, and the spectra of the signal and idler are almost equal. Figure \ref{HOM} (i) is obtained by using a pump bandwidth of $\Delta\lambda$ = 11 nm, and a crystal length L  of 100 mm. The spectra of the signal and idler have the same FWHM of 54.64 nm. The HOM interference from two independent signal or idler sources manifests the same performance with the FWHM of 12.24 ps and visibility of 82.33\%, which is much lower than the GVM$_1$ and GVM$_2$ case. In the case of two-fold HOM interference, the visibility is 100\% and the FWHM of the HOM pattern is 2.20 ps for Fig.\ref{HOM} (l).

\section{Discussion}\label{sec4}

We summarize the result of all the BPM crystals in Fig. \ref{BPM-result}. 
The left vertical axis of the figure denotes the GVM condition and the corresponding  PMF angle $\theta_{PMF}$. The right vertical axis shows the maximal purity. The horizontal axis shows a wavelength range from 0 to 12 $\mu$m. Most of the crystals are located on the MIR band, from 2 $\mu$m to 12 $\mu$m. There are three cases on the NIR band. 
We also conclude the results of QPM crystals in Figure \ref{QPM-result}, which shows the down-converted wavelength, the poling period, and the maximal purity for all the results we calculated above.

It is important to discuss the detection of the single photons in the MIR region. Recent work shows that superconducting nanowire single-photon detectors (SNSPD) which have the best performance (98\%) in the NIR band \cite{Reddy2020}, while having a detection efficiency at MIR band of 70\% at 2 $\mu$m \cite{Chang2022PRJ}, 40\% at 2.5 $\mu$m and 10\% at 3 $\mu$m \cite{Zolotov2017},  1.64\% for free-space communication \cite{Bellei2016OE}; upconversion detectors module combine with SAPD method demonstrates the efficiency of 6.5\% in room temperature \cite{Mancinelli2017NC}; semiconductor photodiodes likes Cd admixture, graphene, black arsenic phosphorus, black phosphorus carbide, tellurene, PtSe$_2$ and PdSe$_2$ are good candidates for wide detection range \cite{Long2018}. See a recent review about MIR single-Photon detection in \cite{Russo2020}. In the future, developing new material for SNSPD and a more effective nonlinear process of upconversion for MIR detection will be promising. 


Expect from the 22 crystals we discussed above, we still find 6 kinds of new crystals  in the MIR band: BGS, BGSe, BGGS, BGGSe, BGSS, and BGSSe \cite{Li2015JMCC, Zhai2017, He2019OE, Kato2020OL, Elu2020OL}. They can be written as BaGa$_4$$X$$_7$ ($X$=S, Se) and BaGa$_2$$M$$X$$_6$($M$ = Si, Ge; $X$=S, Se). Since the d$_{eff}$ calculated and phase-matched method of these crystals is complex \cite{Boursier2015OL, Boursier2016OL, Kato2017AO, Kato2018AO, Liu2022}, we did not discuss them in this work.
The d$_{eff}$ for BGS and BGSe has been investigated in \cite{Badikov2011}. The d$_{33}$ for BGGS, BGGSe, BGSS, and BGSSe is $-$12.0, $-$23.0, 8.4, and 12.3 pm/V, respectively \cite{Yin2012DT}. Moreover, the doping method can be utilized as a degree of freedom to manipulate the single-photon state at the MIR range \cite{Wei2021}.

For the GVM$_3$ condition, the purity can be further improved from 0.82 to near 1 using the custom poling crystal scheme, for example by machine learning method or metaheuristic algorithm \cite{ Cui2019PRApl, Cai2022}. 


%

\section{Conclusion}
In conclusion, we have theoretically investigated 22 nonlinear optical crystals for MIR photon generation. The down-converted photons wavelength range is from 1298 nm (1224 nm) to 11650 nm (7944 nm) for the BPM (QPM) crystals.
The corresponding purity for three kinds of GVM conditions are around 0.97, 0.97, and 0.82, respectively. 
The wavelength nondegenerated condition, the 4-fold HOM interference, and the  2-fold  HOM interference are calculated in detail.
This study may be helpful to the study of quantum communication, quantum imaging, and quantum metrology at MIR range.

\section*{Acknowledgments}
We thank Prof. Keiichi Edamatsu for the helpful discussions.
This work is supported by the National Natural Science Foundations of China (Grant Nos.12074299, 91836102, 11704290), the Guangdong Provincial Key Laboratory (Grant No. GKLQSE202102), the Natural Science Foundation of Hubei Province (2022CFA039), and the JST SPRING (Grant Number JPMJSP2114).

\bibliography{MIRbib}

\end{document}